\documentclass{ecai} 

\usepackage{latexsym}
\usepackage{amssymb}
\usepackage{amsmath}
\usepackage{amsthm}
\usepackage{booktabs}
\usepackage{enumitem}
\usepackage{graphicx}
\usepackage{color}
\usepackage{algorithm}
\usepackage{algpseudocode}
\usepackage{multirow, makecell}

\newcommand{\partitle}[1]{\smallskip \noindent \textbf{#1.}}

\newcommand{\projectname}{{\textsc{\sf SecPE}}}

\newcommand{\sss}{\mathbf{s}}
\newcommand{\yyy}{\mathbf{y}}
\newcommand{\zzz}{\mathbf{z}}

\newcommand{\BibTeX}{B\kern-.05em{\sc i\kern-.025em b}\kern-.08em\TeX}

\begin{document}


\begin{frontmatter}


\paperid{123} 


\title{\projectname: Secure Prompt Ensembling for Private and Robust Large~Language~Models}


\author[A]{\fnms{Jiawen}~\snm{Zhang}\footnote{Equal contribution.}}
\author[A]{\fnms{Kejia}~\snm{Chen}\footnotemark}
\author[A]{\fnms{Zunlei}~\snm{Feng}\thanks{Corresponding Author. Email: zunleifeng@zju.edu.cn}}
\author[A]{\fnms{Jian}~\snm{Lou}\thanks{Corresponding Author. Email: jian.lou@zju.edu.cn}}
\author[A]{\fnms{Mingli}~\snm{Song}}
\author[A,B]{\fnms{Jian}~\snm{Liu}}
\author[A,B]{\fnms{Xiaohu}~\snm{Yang}} 

\address[A]{State Key Laboratory of Blockchain and Data Security, Zhejiang University}
\address[B]{Hangzhou High-Tech Zone (Binjiang) Blockchain and Data Security Research Institute}


\begin{abstract}
With the growing popularity of LLMs among the general public users, privacy-preserving and adversarial robustness have become two pressing demands for LLM-based services, which have largely been pursued separately but rarely jointly. In this paper, to the best of our knowledge, we are among the first attempts towards robust and private LLM inference by tightly integrating two disconnected fields: private inference and prompt ensembling. The former protects users' privacy by encrypting inference data transmitted and processed by LLMs, while the latter enhances adversarial robustness by yielding an aggregated output from multiple prompted LLM responses. Although widely recognized as effective individually, private inference for prompt ensembling together entails new challenges that render the naive combination of existing techniques inefficient. 

To overcome the hurdles, we propose \projectname, which designs efficient fully homomorphic encryption (FHE) counterparts for the core algorithmic building blocks of prompt ensembling. We conduct extensive experiments on 8 tasks to evaluate the accuracy, robustness, and efficiency of \projectname. The results show that \projectname ~maintains high clean accuracy and offers better robustness at the expense of merely $2.5\%$ efficiency overhead compared to baseline private inference methods, indicating a satisfactory ``accuracy-robustness-efficiency'' tradeoff. For the efficiency of the encrypted \textsc{Argmax} operation that incurs major slowdown for prompt ensembling, \projectname ~is 35.4 times faster than the state-of-the-art peers, which can be of independent interest beyond this work.
\end{abstract}

\end{frontmatter}


\section{Introduction}
Large language models (LLMs) have garnered a meteoric rise in popularity among general public users due to their remarkable performance across myriad natural language processing (NLP) tasks \cite{xu2019bert,yang2019end}. LLMs are oftentimes deployed by service providers in the form of Machine Learning as a Service (MLaaS) \cite{yang2019simple,raffel2020exploring}, whereby users can conveniently exploit the full potential of LLM by submitting their inference data, prepended by specific prompts from prompt learning techniques \cite{li2023exploring}, to obtain high-performing LLM outputs tailored to their downstream tasks. Accompanying this widespread adoption, there arise privacy and robustness concerns for LLMs \cite{juvekar2018gazelle}.

\partitle{Privacy concerns and private inference} 
On the privacy aspect, users' inference data can inadvertently reveal sensitive information if transmitted and processed by the LLM service provider in plaintext~\cite{yang2019simple,raffel2020exploring},  risking identification and privacy breaches. Additionally, the user-submitted prompts can be valuable intellectual property and also raise privacy concerns. As a result, both inference data and user-side prompts demand privacy-preserving measures \cite{juvekar2018gazelle, zhang2024secure}. Among the many attempts to avoid submitting raw data for LLM inference, private inference offers very strict privacy protection by allowing inference to be conducted on encrypted data. For instance, Fully Homomorphic Encryption (FHE) allows rich computations (covering most operations needed in LLM inference) on encrypted data without exposing sensitive information~\cite{gentry2009fully}. By encrypting inputs using FHE, only encrypted predictions are sent to the server, ensuring privacy throughout the process. As legal and societal pressures mount, service providers' adoption of such privacy-preserving technologies has received increasing research attention.

\partitle{Robustness concern and prompt ensembling} 
On the robustness aspect, it is well-recognized that the output of LLMs can be manipulated by subtle yet deliberate changes in the inference sample or the prompt \cite{wang2024towards}. There has been a growing focus on enhancing the robustness of LLMs, especially in safety-critical downstream application areas. Various methods have been proposed, ranging from more advanced (and sophisticated) to simple methods \cite{dvornik2019diversity}. One representative method from the latter category follows the idea of prompt ensembling \cite{schick2020exploiting}, which involves making multiple inferences for a single inference data and providing the aggregated result as the final prediction. 

\partitle{This study} 
The current research efforts on safeguarding privacy and robustness during LLM inference are largely explored separately. Driven by the simultaneous demands from both privacy and robustness aspects, we envision that these two aspects should be pursued jointly. Among the first attempts toward mitigating both concerns of LLMs jointly, we investigate the potential to achieve private and robust LLM inference through tight integration of private inference and prompt ensemble. We focus on these two techniques due to their effectiveness in addressing their respective concerns. In particular, we note that while there may be more advanced techniques for enhancing robustness than prompt ensembling, achieving a balance between robustness and efficiency within the private inference workflow of the simpler prompt ensembling method already poses significant challenges. That is, naive application of existing private inference methods for prompt ensembling entails great efficiency overhead. The crux of efficient private inference for prompt ensembling is that the aggregation operation introduced by prompt ensembling, albeit simple and efficient in plaintext computation, requires prohibitive computation in the ciphertext. 

\begin{figure}[tbp]
    \centering
    \includegraphics[width=3.2in]{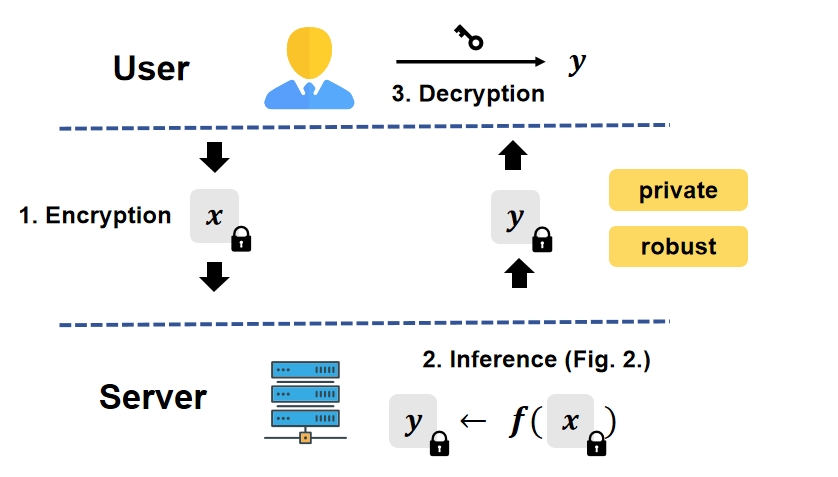}
    \caption{A high-level overview of \projectname ~for private and robust LLM inference in FHE-based MLaaS.}
    \label{fig:overview}
\end{figure}

To overcome the inefficiency challenges, we propose \projectname: a new secure prompt ensembling method for private and robust LLM inference. As illustrated in Figure \ref{fig:overview}, \projectname ~allows user to encrypt their inference data and prompts before transmitting them to the LLM server for inference. The inference results from the LLM server are aggregated from multiple prompted responses and transmitted back to the user in ciphertext format, which can be decrypted only by the user’s private key. The encrypted aggregation operation heavily relies on efficient computation of \textsc{Argmax}, which is unfortunately not readily supported by the common homomorphic primitives like the RNS-CKKS FHE scheme \cite{lee2022low}. Lying at the design core of \projectname ~is a new efficient private aggregation algorithm to be presented in Algorithm \ref{alg:argmax}, which resorts to an efficient approximation of \textsc{Argmax} to circumvent this efficiency bottleneck. We conduct extensive experiments to test the accuracy, robustness, and efficiency of \projectname ~across 6 tasks from GLUE/AdvGLUE and 2 tasks from mathematical reasoning data sets. We also extend our \projectname ~ to Text-Image Models and test the accuracy across 7 tasks from CIFAR, ImageNet, and their variants. The results show that \projectname ~is capable of maintaining both high utility and robustness while providing privacy protection.

The main contributions of this paper are summarized as follows:
\begin{itemize}[leftmargin=*]
    \item To the best of our knowledge, we are among the first to jointly study the privacy and robustness concerns of LLM inference, which become increasingly pressing considering the growing deployment of LLM-based services.
    \item We propose \projectname ~to achieve private and robust LLM inference, which devises new secure primitives tailor-made for prompt ensembling to strike a satisfactory ``accuracy-robustness-efficiency'' tradeoff.
    \item We conduct extensive experiments on 15 tasks from 4 popular benchmarks to corroborate the superior performance of \projectname ~against baseline methods. 
\end{itemize}

\section{Background}
\subsection{Privacy Issues of LLMs}
LLMs such as the GPT have revolutionized natural language processing and understanding with human-level proficiency~\cite{kenton2019bert,brown2020language}. 
However, with their increasing deployment in MLaaS by service providers and growing popularity among the general public users, there arise aggravating privacy concerns.
In the typical MLaaS serving setting, users submit inference data to the remote server hosting a proprietary model and receive predictions in return. Users therefore have privacy concerns about their inference data that, despite being sensitive or even confidential, are transmitted and processed in plaintext by the MLaaS service provider ~\cite{shen2007privacy}. This issue has even led to ChatGPT being temporarily banned in Italy~\cite{mauran2023whoops, Italyorders}. Recognizing this pressing privacy concern, existing works introduce various means to avoid direct transmission and processing inference data in plain text form. 

Private inference emerges as a viable solution, promising to reconcile the need for high-performant inference data processing with strict privacy requirements~\cite{srinivasan2019delphi,hao2022iron,pang2023bolt}. 
Private inference provides a way to guarantee the privacy and confidentiality of both the inference data and the proprietary LLM. It ensures that data is not transmitted or processed in plaintext but as ciphertext, thereby safeguarding sensitive details about the server's model weights and the user's inputs from disclosure. 
While private inference has significant applications in computer vision and image processing~\cite{zeng2023mpcvit}, its use in LLMs is nascent. Notably, the integration of private inference in prompt learning settings and prompt ensembles remains an under-explored area, presenting a frontier yet to be ventured into the field.

By pursuing private inference tailored for prompt ensemble learning, we aim to bridge the gap between utility, robustness, and privacy, thereby realizing the benefits of prompted LLMs without compromising user trust and data integrity.

\subsection{Fully Homomorphic Encryption}

\begin{figure*}[t]
    \centering
    \includegraphics[width=6.6in]{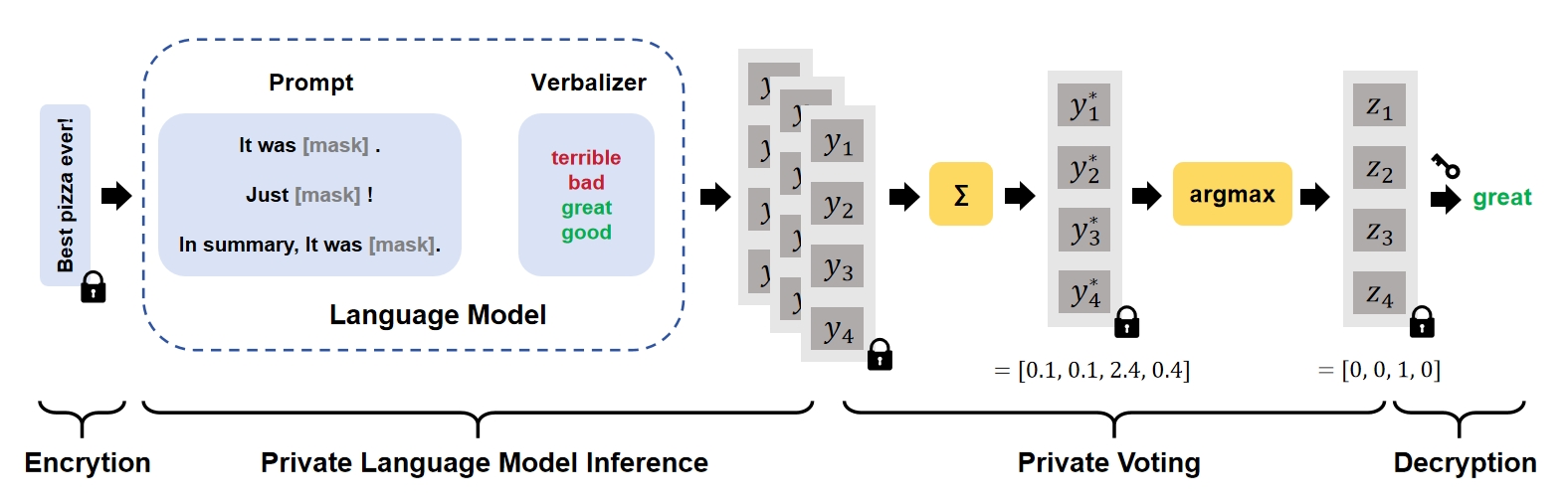}
    \caption{An illustration of {\sc secPE}, which enables homomorphically encrypted LLM inference with guarantees.}
    \label{fig:secpe}
\end{figure*}

The FHE scheme used in this paper is the full {\em residue number system} (RNS) variant of Cheon-Kim-Kim-Song (CKKS)~\cite{cheon2019full}.
RNS-CKKS is a {\em leveled} FHE, which can support computations up to a multiplicative depth $L$. 
Both the plaintexts and ciphertexts of RNS-CKKS are elements in a polynomial ring:  
$$\mathcal{R}_{Q}=\mathbb{Z}_{Q}[X]/(X^{N}+1)$$
where $Q = \Pi_{i=0}^L q_i$ with distinct primes $q_i$.
Once a ciphertext's level becomes too low, a {\em bootstrapping} operation is required to refresh it to a higher level, enabling more computations.
In a nutshell, bootstrapping homomorphically evaluates the decryption circuit and raises the modulus from $q_0$ to $q_L$ by leveraging the isomorphism $\mathcal{R}_{q_0} \cong \mathcal{R}_{q_0} \times \mathcal{R}_{q_1} \times \cdot\cdot\cdot \times \mathcal{R}_{q_L}$~\cite{bossuat2021efficient}. 
Suppose the bootstrapping consumes $K$ levels, then a fresh ciphertext can support $L-K$ levels of computations.

RNS-CKKS supports {\em single instruction multiple data} (SIMD), which enables encrypting a vector with $N$ elements into a single ciphertext and processing these encrypted elements in a batch without introducing any extra cost. Below, we summarize the homomorphic operations used in this paper:

\begin{itemize}
    \item $a \oplus b$. The addition takes two SIMD ciphertexts $a$ and $b$; outputs  $[a_0+b_0, a_1+b_1...,a_{N-1}+b_{N-1}]$.
    \item $a  \ominus  b$. The subtraction takes two SIMD ciphertexts $a$ and $b$; outputs  $[a_0-b_0, a_1-b_1...,a_{N-1}-b_{N-1}]$.
    \item $a \otimes b$. The multiplication takes two SIMD ciphertexts $a$ and $b$; outputs  $[a_0 \times b_0, a_1 \times b_1...,a_{N-1} \times b_{N-1}]$.
    \item $RotL(a, s)$. The left-rotation takes one SIMD ciphertext $a$ and an integer $s$; left-rotates the vector by $s$ slots.
    \item $RotR(a, s)$. The right-rotation takes one SIMD ciphertext $a$ and an integer $s$; right-rotates the vector by $s$ slots.

\end{itemize}

\subsection{Prompt Ensembling for Robust LLMs}
The brittleness of LLMs to slight input modifications often leads to varied/inaccurate and sometimes even malicious/harmful outputs, highlighting the essential need for enhanced robustness for LLMs~\cite{schick2020automatically}. Robustness in this context refers to LLM's ability to provide consistent predictions regardless of slight changes to the inference data, aiming for more predictable and stable responses. 

Building on the success of prompt learning, prompt ensemble learning \cite{allingham2023simple} demonstrates the potential to offer efficient, effective, and robust predictions. Prompt ensemble utilizes a series of prompts to allow for the aggregation of multiple responses for the same inference data, leading to more robust predictions. 

Prompt ensembling, in which the masked language model $\mathcal{L}$ is directly tasked with "auto-completing" natural language prompts. For instance, for the inference data $x_{in}$, the template into which the inference data is inserted that $x_{\text{prompt}}$ = “It was \texttt{MASK}” is concatenated (i.e., $x_i=x_{in}$ $\oplus$ $x_{\text{prompt}}$), The prompt typically includes one or more masked tokens $\texttt{[MASK]}$ that the model $\mathcal{L}$ is expected to fill in, making it a structured query that directs the model's response.

The single output refers to the model's prediction for each prompt, drawing on the context of the prompt and input data present, like determining the sentiment of a movie review. When multiple prompts or input variations are used to obtain a range of model responses, the aggregated output synthesizes these individual outputs to derive a more robust or accurate prediction. This aggregation could involve combining the model's responses to enhance prediction reliability or accuracy, especially in tasks where nuanced understanding or multiple aspects of the input data are considered.

For NLP tasks, suppose there are $m$ prompt templates, the verifier takes a question and a candidate reasoning path as input and outputs the probability that the reasoning path leads to the correct answer~\cite{li-etal-2023-making}. 
$$
y^*=\textsc{Argmax}(\sum_{i=1}^m f(x_{in} \oplus x_{\text{prompt}_i}))
$$
where $f(\cdot)$ is the probability produced by the verifier's model $\mathcal{L}$.

In addition to LLM, prompt ensembling is also widely used in text-image models. For zero-shot image classification tasks, suppose there are $m$ prompt templates, prompt ensembling as
proposed in CLIP \cite{radford2021learning} generalizes: 
$$
y^*=\textsc{Argmax}(\sum_{i=1}^m (I(x_{in}) \cdot T(x_{\text{prompt}_i})))
$$
where $I$ is the image encoder and $T$ is a text encoder.

\section{Proposed Method: \projectname}

We propose a new private inference framework tailor-made for the prompt ensembling. Private inference for prompt ensembling raises a critical, unaddressed issue: the challenge of integrating private contextual inference. Incorporating privacy-preserving mechanisms into prompt ensembles remains a significant and complex challenge, despite progress in leveraging prompt-based learning to improve model effectiveness in downstream tasks. Our work aims to break new ground by developing a comprehensive framework that not only improves model performance through optimized prompt selection but also prioritizes the integration of robust privacy safeguards. 

\subsection{\projectname ~Framework}
We give an illustration of \projectname ~in Fig \ref{fig:secpe}, the overall process is divided into the following four steps:
\begin{enumerate}[leftmargin=*]
    \item \textbf{Encryption.} User encrypts $m$ inputs $x_i = x_{in} \oplus x_{\text{prompt}}, i \in [1, m]$ using FHE and sends them to the server, where $m$ is the number of prompt templates.
    \item \textbf{Private Language Model Inference.} Server uses the language model $\mathcal{L}$ classifying $m$ inputs into one of $n$ classes, $n$ is the number of labels. the inputs are propagated through $\mathcal{L}$ utilizing the homomorphic operations of the FHE scheme to obtain $m$ encrypted logits $y_i, i \in [1, m]$.
    \item \textbf{Private Voting.} Server aggregates the encrypted logits $y^* \gets \sum_{i=1}^m y_i$ and then evaluates \textsc{Argmax} function in FHE. In particular,
    this step transforms the logit vector $y^*$ into a one-hot vector $z$. Then the server sends $z$ to the User.
    \item \textbf{Decryption.} User decrypts $z$ with its secret key, where the single non-zero entry represents the index of the classification label.
\end{enumerate}

As illustrated in the preceding workflow of \projectname, Steps 1 and 4 pertain to fundamental FHE encryption and decryption operations. Step 2 has been implemented across numerous recent works, including in \cite{hao2022iron, pang2023bolt}. These three steps are orthogonal to the efficiency designs of prompt ensembling. The primary obstacle lies in leveraging FHE to access the \textsc{Argmax} operation in Step 3. 

It is important to note that private voting can only be calculated by the server in ciphertext and cannot be handed over to the user in plaintext. This is because numerous works, including those by\cite{shokri2017membership, ye2022enhanced, yan2022membership}, have designed membership inference attacks based on the class probability distribution of the prediction vector. It is for this reason that the output of the final layer, commonly referred to as the logits, is generally considered to represent the raw confidence ratings associated with the predictions. These ratings are selected using the \textsc{Argmax} processing, whereby the one with the highest probability is selected from the available ratings. It is important to note that the simple act of returning these logits without the \textsc{Argmax} process carries the risk of exposing more information about the underlying data and the decisions made by the model, potentially resulting in privacy leakage.


The fact that FHE does not permit control flow evaluation (e.g., branching) and that ciphertext comparison (e.g., inequality checking) is not directly supported by the homomorphic primitives of the RNS-CKKS FHE scheme means that we cannot implement the \textsc{Argmax} algorithm in a canonical manner. Instead, we are seeking an efficient approximation to circumvent the efficiency bottleneck that is introduced by prompt ensembling.

\subsection{Efficient Private Inference for Prompt Ensembling}
As mentioned above, the design core of efficient private inference for prompt ensembling lies at the private aggregation operator, i.e., the \textsc{Argmax} operation. 

Therefore, our goal is to approximate the following function on an RNS-CKKS ciphertext logit vector:
\begin{equation}\label{eq:argmax}
    [y_1, ..., y_n, 0^{N-n}] \rightarrow [z_1,  ..., z_n, \#^{N-n}],
\end{equation}
where  $z_i=1$ for the index $i$ corresponding to the largest value
among $[y_1, y_2, ..., y_n]$ (and 0 elsewhere). 

The state-of-the-art non-interactive protocol that can achieve this goal is Phoneix~\cite{jovanovic2022private}. Phoenix adopts the idea of bubble sorting to compare each element with adjacent elements by rotating the ciphertext and making a difference with the input:
\begin{align*}
    {\sss_1} &\gets Sign({\yyy} - RotL({\yyy}, 1)) \\
    {\sss_2} &\gets Sign({\yyy} - RotL({\yyy}, 2)) \\
    &... \\
    {\sss_m} &\gets Sign({\yyy} - RotL({\yyy}, m))
\end{align*}

So ${\sss} \gets \sum_{i=1}^{m}{\sss_i}$ counts the comparison result among each input and adjacent elements. Obviously, the value of the maximum element position is $m$, and the values of other positions are less than $m$. After that, through simple linear transformation, ${\zzz}$ can be obtained based on ${\sss}$ (cf. Phoenix \cite{jovanovic2022private} for details).

However, this method requires  $(m+1)$ times $Sign$ operations and $(m+1)$ times ciphertext rotations, which is very inefficient when $m$ is large (e.g. $m=1024$ in CLIP \cite{radford2021learning}). To solve the problem, we innovatively proposed an \textsc{Argmax} evaluation method as:
\begin{equation}\label{eq:new_argmax}
    z_i \gets Sign(y_i - y_{max}) + 1.
\end{equation}

To enable encrypted comparisons, we leverage the polynomial approximation of the sign function:

\begin{equation}
    Sign(x)=
    \begin{cases}
          -1 \quad & \ -1 \leq x \leq -2^{-\alpha} \\
          0 &\quad\quad  x = 0 \\
          1 &\quad  2^{-\alpha} \leq x \leq 1 \\
     \end{cases}
\end{equation}

The approximation involves a composition of polynomials:

\begin{equation}
    Sign(x) = f^{d_f}( g^{d_g}(x))
\end{equation}

where $f(), \; g()$ are two polynomials and $d_f, \; d_g$ are the number of repetitions for them. In our implementation, both $f()$ and $g()$ are 9-degree polynomials; we set $\alpha = 12, d_f=2, d_g=2$, so the max error bound is less than $10^{-4}$. To reduce the multiplicative depth, we evaluate the polynomials using the Baby-Step-Giant-Step algorithm~\cite{han2020better}. 

Before proceeding, we comment on the basic input requirement of $Sign(x)$, namely
that its inputs are in $[-1, 1]$. Suppose the inputs $x_i \in [D_{min}, D_{max}]$, to ensure this requirement, for those inputs that need to be different from each other, we need to normalize $\hat{x_i} \in [0,1]$:

\begin{equation}
    \hat{x_i} = \frac{x_i-D_{min}}{D_{max} - D_{min}},
\end{equation}

meaning that for all $i \neq j$, $\hat{x_i} - \hat{x_j} \in [-1, 1]$, satisfying the requirement in Algo.\ref{alg:argmax}

In order to get $x_{max}$, with the help of the $Sign$ function, we can calculate the maximum value of $a$ and $b$ by:

\begin{equation}\label{eq:max}
    Max(a,b) = \frac{a+b}{2} + \frac{a-b}{2} \cdot Sign(a-b).
\end{equation}

Then, the selection vector can be easily computed as described in Algorithm~\ref{alg:argmax}. 

\begin{algorithm}[t]
\caption{Private \textsc{Argmax} on RNS-CKKS}\label{alg:argmax}
\begin{algorithmic}[1]
\Function{\textsc{Argmax}}{$y$}
\State $y \gets y \oplus RotR(y, n)$
\State $y_{max} \gets QuickMax(y)$
\State $y \gets y \ominus y_{max}$
\State $z \gets Sign(y)$
\State $z \gets z \oplus 1$ 
\State \Return $z$
\EndFunction
\Function{$QuickMax$}{$y$}
\State $l \gets \log_2n$
\For{$i=0$ to $\log n-1$}
  \State $r \gets RotL(y, 2^i)$
  \State $r \gets Max(r, y)$
  \State $y \gets r$
\EndFor
\State \Return $y$
\EndFunction
\end{algorithmic}
\end{algorithm}

\begin{figure}[!h]
    \centering
    \includegraphics[width=3in]{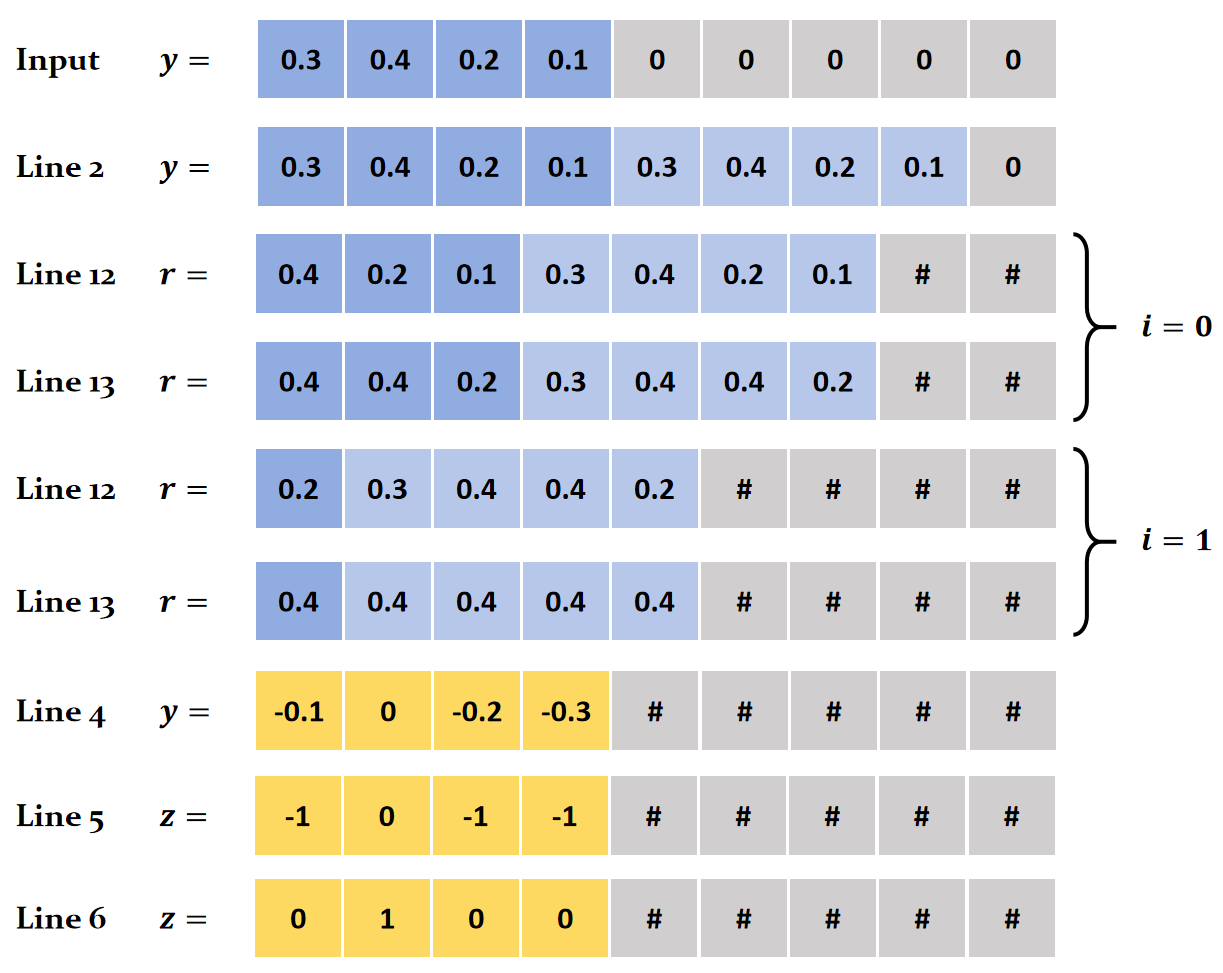}
    \caption{Example run of Algorithm \ref{alg:argmax}.}
    \label{fig:argmax}
\end{figure}

In Fig. \ref{fig:argmax}, we illustrate how Alg. \ref{alg:argmax} processes a toy example. The algorithm first duplicates the logits (Line 2), then uses $QuickMax$ to get the maximum value of $[y_1, y_2, ..., y_n]$. Unlike phoenix \cite{jovanovic2022private}, we do not rotate only one step at a time, but rotate $2^i, \;i \in [0, \log n-1]$ steps each time, which greatly reduces our number of rotations and the number of $Sign$ operations. This technique can be applied to all associative operations, such as sum, maximum or minimum, etc.

Note that we need to use $2n$ slots to calculate the $\textsc{Argmax}$ of an input of length $n$. Since the total number of slots of the ciphertext polynomial is $N \gg n$, we can batch process the $\textsc{Argmax}$ of $\frac{N}{2n}$ inputs in parallel in a polynomial ciphertext. For example, if the SIMD slot is 32768, and the input length is 256, we can batch 64 inputs in parallel.

\begin{table*}[h!]
\centering
{\small
\setlength{\tabcolsep}{5pt}
\begin{tabular}{ccccccccc}
\hline\\[-2ex]
\textbf{Method} & \textbf{Prompt} & \textbf{Setting} & \textbf{SST-2} & \textbf{QQP}  & \textbf{MNLI-m}  & \textbf{MNLI-mm}  & \textbf{RTE}  & \textbf{QNLI}
\\[0.5ex]\hline\\[-1.5ex]
\multirowcell{2}{LM-BFF (Plaintext)} & \multirowcell{2}{Single}  & Cln & 94.0 & 80.1 & 76.7 & 78.3 &  78.1 & 81.4 \\ && Adv &  54.1 & 46.2 & 47.1 & 40.1 & 58.8 & 61.5 \\[0.5ex]\hline\\[-1.5ex]
\multirowcell{2}{LM-BFF (Ciphertext)} & \multirowcell{2}{Single} & Cln & 93.7 & 79.2 & 76.0 & 77.6 &  77.5 & 81.0 \\ && Adv &  53.8 & 46.1 & 46.4 & 39.5 & 58.2 & 61.1 \\[0.5ex]\hline\\[-1.5ex]
\multirowcell{2}{PET (Plaintext)} & \multirowcell{2}{Ensemble} & Cln & 93.4 & 73.7 & 74.6 & 75.7 &  74.2 & 84.6 \\ && Adv &  61.7 & 59.3 & 55.6 & 44.8 & 54.0 & 67.9 \\[0.5ex]\hline\\[-1.5ex]
\multirowcell{2}{\textbf{\projectname}} & \multirowcell{2}{Ensemble} & Cln & 93.0 & 73.1 & 73.2 & 74.7 &  72.2 & 81.1 \\ && Adv & 61.3 &  59.3 & 55.4 & 43.9 & 53.2 & 66.8 \\[0.5ex]\hline\\[-1.5ex]
\end{tabular}
}
\vspace{-1em}
\caption{Performance comparison on GLUE (Cln) and Adversarial GLUE (Adv) benchmarks. We report the average and standard deviation in the accuracy values of 5 different runs.}
 \label{table:benchmark}
 \vspace{-1em}
\end{table*}
\section{Experiments}

\subsection{Experimental setup}

\partitle{Tasks and Datasets}

In the experiments, we utilize 8 tasks from popular benchmarks to thoroughly evaluate the utility, robustness, and efficiency of {\sf SecPE}.

\noindent\underline{I) Benign NLP tasks.} We evaluate {\sf SecPE} on six tasks from the \textbf{GLUE} benchmark. In detail, the evaluated tasks are (1) SST-2; (2) QQP; (3) MNLI-matched; (4) MNLI-mismatched, (5) RTE, and (6) QNLI—range, which range from sentiment analysis to question answering, diversifying in different inference data formats from sentences to pairs of sentences.

\noindent\underline{II) Adversarial NLP tasks.} We evaluate the robustness of {\sf SecPE} on six adversarial tasks in the Adversarial-GLUE (\textbf{AdvGLUE}) benchmark~\cite{wang2021adversarial}, which are adversarial counterparts to the above benign GLUE tasks.
The AdvGLUE benchmark is enriched with task-specific adversarial examples generated by 14 different textual attack methods, coming from different adversarial perturbation strategies including word-level, sentence-level, and human-generated. Recognizing the potential problem of invalid adversarial constructs identified by Wang et al.~\cite{wang2021adversarial}, where up to 90\% of automatically generated examples may be flawed, we also incorporate human validation. This step allows for a more accurate and robust evaluation of {\sf SecPE} by ensuring that the adversarial examples in our benchmark are legitimate and that the perturbations maintain the integrity of the original task.

\noindent\underline{III) Arithmetic reasoning tasks.} We evaluate the self-consistency of \projectname~ on ~two arithmetic reasoning benchmarks: GSM8K~\cite{cobbe2021training} and MultiArith~\cite{roy2016solving}. GSM8K contains grade-school-level mathematical word problems requiring models to perform complex arithmetic reasoning and multi-step calculations. MultiArith contains multiple arithmetic operations within a single problem, testing a model's ability to comprehend and execute a sequence of calculations, reflecting the complexity of mathematical reasoning needed for higher accuracy in various problem-solving contexts. 

\noindent\underline{IV) Zero-shot image classification tasks.} We also evaluate {\sf SecPE} on ImageNet and its variant test sets ImageNet-R , ImageNet-A , ImageNet-Sketch and ImageNet-V2. We also evaluate on CIFAR10 and CIFAR100, which
are fine-grained classification datasets.

\begin{table}[tbp]\scriptsize
\centering
\begin{tabular}{@{}l|l|l@{}}
\toprule
\textbf{Task}\ & \textbf{Template} & \textbf{Verbalizer} \\ \midrule
\multirowcell{3}{\textbf{SST-2}} & It was $\texttt{[MASK]}$ . $<S_1>$ & bad / good \\
      & $<S_1>$ . All in all, it was $\texttt{[MASK]}$. & bad / good \\
      & Just $\texttt{[MASK]}$! $<S_1>$ & bad / good \\
      & In summary, the movie was $\texttt{[MASK]}$. & bad / good \\ \addlinespace \midrule
\multirowcell{3}{\textbf{QQP}}   & $<S_1>$ $\texttt{[MASK]}$, $<S_2>$ & No / Yes \\
      & $<S_1>$ $\texttt{[MASK]}$, I want to know $<S_2>$ & No / Yes \\
      & $<S_1>$ $\texttt{[MASK]}$, but $<S_2>$ & No / Yes \\
      & $<S_1>$ $\texttt{[MASK]}$, please, $<S_2>$ & No / Yes \\ \addlinespace \midrule
\multirowcell{3}{\textbf{MNLI}}  & $<S_1>$ ? $\texttt{[MASK]}$, $<S_2>$ & Wrong/Right/Maybe \\
      & $<S_1>$ ? $\texttt{[MASK]}$, $<S_2>$" & No/Yes/Maybe \\
      & $<S_1>$ ? $\texttt{[MASK]}$, $<S_2>$ & Wrong/Right/Maybe \\ \addlinespace \midrule
\multirowcell{3}{\textbf{RTE}}   & "$<S_2>$ ? $\texttt{[MASK]}$, $<S_1>$" & No/Yes \\
      & "$<S_2>$ ? $\texttt{[MASK]}$, $<S_1>$ & No/Yes \\
      & "$<S_1>$ ? $\texttt{[MASK]}$. $<S_2>$ & No/Yes \\ \addlinespace \midrule
\multirowcell{3}{\textbf{QNLI}}  & $<S_1>$ ? $\texttt{[MASK]}$, $<S_2>$ & No/Yes \\
      & $<S_1>$ ? $\texttt{[MASK]}$, $<S_2>$ & Wrong/Right \\
      & "$<S_1>$ ? $\texttt{[MASK]}$, $<S_2>$" & Wrong/Right No/Yes \\ \bottomrule
\end{tabular}
\caption{Manual template and verbalizer pairs. $<S_1>$ and $<S_2>$ are the input sentences.}
\label{table:template}
\end{table}

\partitle{Private Inference Implementation}
We develop encryption functions with C++ and integrate the  SEAL library for RNS-CKKS homomorphic encryption. To improve performance on Intel CPUs, we include HEXL acceleration. Our configuration adheres to homomorphic encryption standards, setting the polynomial degree to \(N=2^{16}\) and the ciphertext modulus to 1763 bits for 128-bit security. We set a multiplicative depth of \(L=35\) and a bootstrapping depth of \(K=14\), resulting in an effective multiplicative depth of \(21\).

\subsection{Evaluation Results on GLUE and Adversarial GLUE Tasks}

For tasks within the GLUE and AdvGLUE benchmarks, we use the ALBERT-XXLarge-v2 model \cite{lan2019albert} to generate different contextual representations. This combined text is fed into the model to obtain the language model results. This method allows us to assess the relationship between questions and their corresponding answers, taking advantage of the model's pre-trained capabilities.

In Table \ref{table:benchmark}, we present evaluation results on GLUE and AdvGLUE tasks, reporting metrics F1 score for QQP and accuracy for the other five tasks). BERT is used as the large pre-trained language model. For baselines LM-BFF and PET, we implement the same private ALBERT-xxlarge-v2 for fair comparison. For the baseline methods, we compare \projectname ~with LM-BFF~\cite{gao-etal-2021-making}, and PET~ \cite{schick-schutze-2021-exploiting}.
\begin{itemize}[leftmargin=*]
    \item \textbf{LM-BFF~\cite{gao-etal-2021-making}:} It involves concatenating the input example, which is modified to follow the prompting template with a \texttt{[MASK]} in place of the verbalizer, with semantically similar examples. During inference, LM-BFF ensembles the predictions made by concatenating the input example with all demonstrations from the few-shot training set.  (i.e., demonstrations) from the few-shot training set. For each test example, we ensemble the predictions over different possible sets of demonstrations. we perform random sampling and subsequent training of LM-BFF for 5 times and 1000 training steps, for each task.
    \item \textbf{PET~\cite{schick-schutze-2021-exploiting}:} It is a simple prompt-based few-shot fine-tuning approach where the training examples are converted into templates, and the \texttt{[MASK]} tokens are used to predict the verbalizer, which indicates the output label. To understand the role of using multiple prompts in robustness, we use PET to fine-tune models with different template-verbalizer pairs and ensemble their predictions during inference. The pairs used for different tasks are listed in Table \ref{table:template}. We train the model on four different sets of manual template-verbalizer pairs for 250 training steps.
\end{itemize}

According to Table \ref{table:benchmark}, we have the following experiment results: 
\begin{itemize}[leftmargin=*]
    \item Compared with prompt ensembles without privacy preservation, \projectname ~exhibits almost no accuracy loss on GELU and AdvGELU benchmarks. This suggests that \projectname ~ is capable of maintaining both high utility and robustness while providing privacy protection.
    \item Compared with the private inference of a single prompt template, \projectname ~has demonstrated better adversarial robustness than LM-BFF(Ciphertext).
\end{itemize}

\subsection{Comparison on Arithmetic Reasoning Tasks}

For reasoning tasks such as MultiArith and GSM8K, we used the GPT-3 model, specifically the code-davinci-001 variant \cite{chen2021evaluating}. This model was chosen for its advanced ability to handle complex language patterns and to generate coherent, contextually relevant text completions. 

The Self Consistency approach employs an array of diverse reasoning pathways, each of which may lead to a different final answer. To identify the optimal answer, we marginalize out the sampled reasoning pathways using a voting verifier (aggregate-then-argmax) as described in \cite{li-etal-2023-making}, thereby determining the most consistent answer in the final answer set.

\begin{figure}[htbp]
    \centering
    \includegraphics[width=3in]{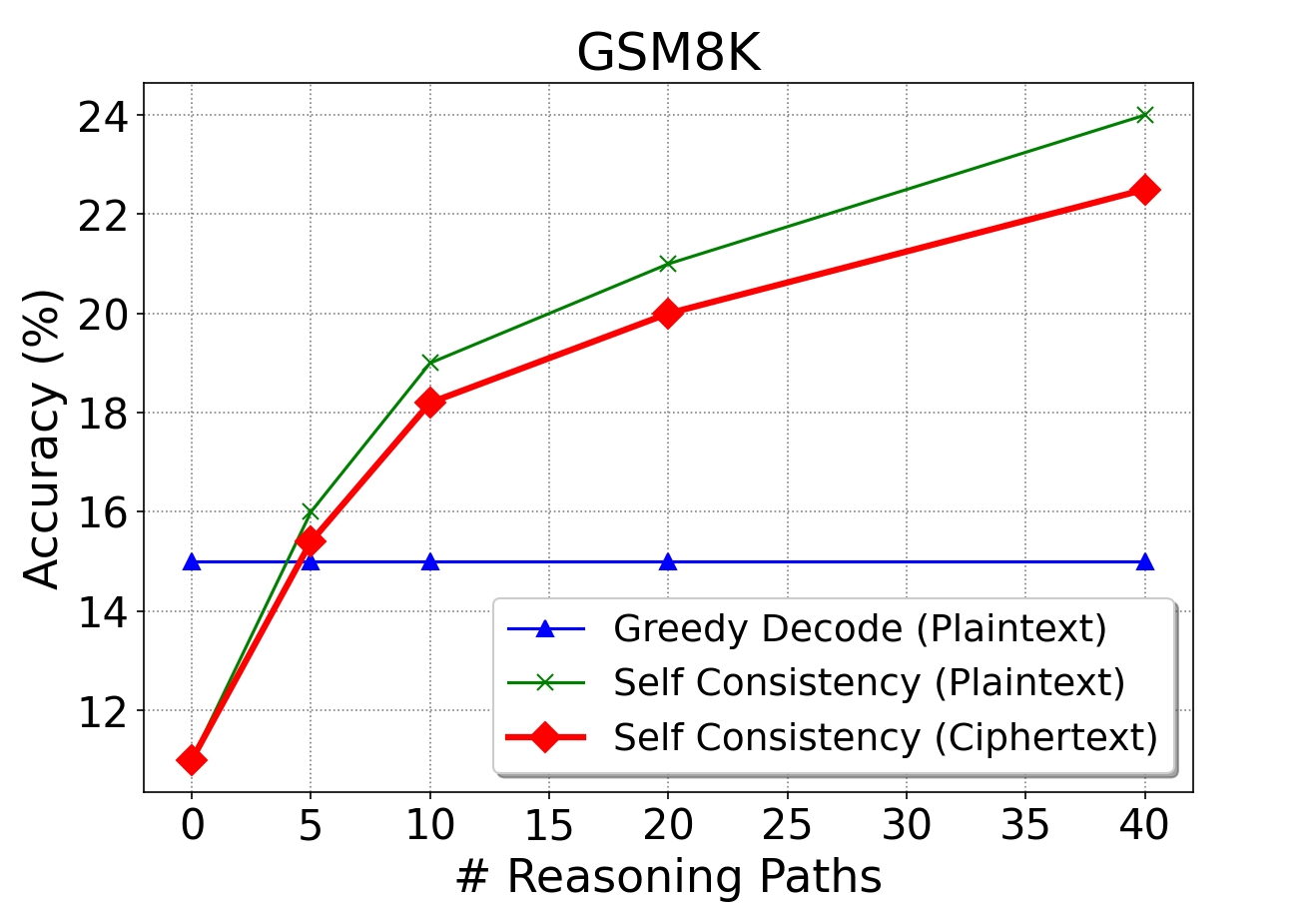}
    \caption{Performance on GSM8K with the different number of reasoning paths.}
    \label{fig:gsm8k}
    \vspace{-1em}
\end{figure}

\begin{figure}[!h]
    \centering
    \includegraphics[width=3in]{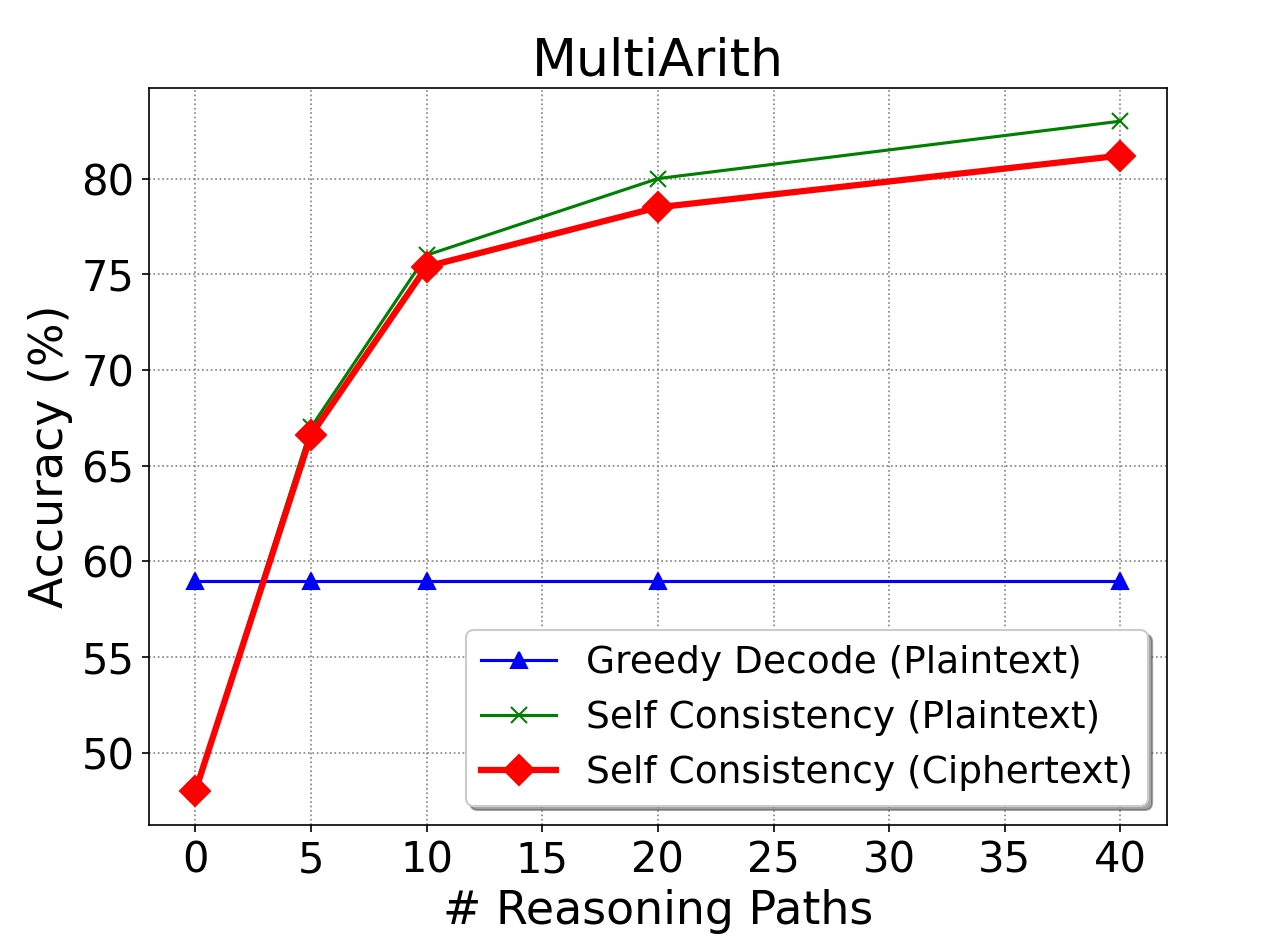}
    \caption{Performance on MultiArith with the different number of reasoning paths.}
    \label{fig:multiarith}
\end{figure}

Under the \projectname ~framework, we have implemented Self Consistency's privacy inference. The baseline to which we are comparing is the chain of reasoning with greedy decoding~\cite{wei2022chain}. The accuracy of the ciphertext inference is similar and much higher than the baseline when compared to the self-consistency inference results under plaintext. Figures \ref{fig:gsm8k} and \ref{fig:multiarith} show the performance on GSM8K and MultiArith with different numbers of inference paths.


\subsection{Comparison on text-image models}
We extend \projectname ~ to Text-Image Models and test the accuracy across 7 tasks from CIFAR, ImageNet, and their variants. We tested zero-shot accuracy with different number of prompts under the CLIP ViT-B/16 model. This model integrates the vision transformer architecture with large language model processing methodologies. This configuration utilizes the "Base" model variant with input patches sized at $16 \times 16$, facilitating the processing of visual data through self-attention mechanisms that are typically reserved for textual data.

We tested with different numbers of prompts:
\begin{itemize}
    \item \textbf{1 prompt.} We used CLIP's most commonly used template "A photo of \texttt{[MASK]}."
    \item \textbf{80 prompts.} We use the set of 80 prompts designed by \cite{radford2021learning} for ImageNet.
    \item \textbf{247 prompts.} We constructed the set of 247 prompts using GPT-4, such as "this is a photo of \texttt{[MASK]}", "A drawing of a \texttt{[MASK]}", etc.
\end{itemize}

\begin{table*}[h!]
\centering
\renewcommand{\arraystretch}{1.2} 
\begin{tabular}{ccccccccc}
\hline
\textbf{Setting} & \textbf{Method} & \textbf{ImageNet} & \textbf{IN-A}  & \textbf{IN-R}  & \textbf{IN-Sketch}  & \textbf{IN-V2}  & \textbf{CIFAR-10} & \textbf{CIFAR-100}
\\[0.5ex]\hline\\[-1.5ex]
\multirowcell{3}{1 prompt} & Plaintext & 66.37 & 47.47 & 73.78 & 45.84 & 60.46 &  96.2 & 83.1 \\ & Phoenix (Ciphertext) &  66.24 & 47.38 & 73.58 & 45.21 & 60.44 & 95.6 & 82.5 \\ & SecPE (Ciphertext) &  66.31 & 47.43 & 73.76 & 45.84 & 60.44 & 95.8 & 82.9

\\[0.5ex]\hline\\[-1.5ex]

\multirowcell{3}{80 prompts} & Plaintext & 67.63 & 49.37 & 77.38 & 46.95 & 61.39 &  96.8 & 84.3 \\ & Phoenix (Ciphertext) &  66.35 & 47.88 & 75.75 & 45.26 & 60.50 &  95.8 & 82.7  \\ & SecPE (Ciphertext) &  67.42 & 49.29 & 77.11 & 46.57 & 61.08 &  96.4 & 83.8 

\\[0.5ex]\hline\\[-1.5ex]

\multirowcell{3}{247 prompts} & Plaintext & 68.60 & 49.63 & 77.62 & 47.99 & 62.21 &  97.5 & 87.9 \\ & Phoenix (Ciphertext) &  66.71 & 48.05 & 75.82 & 46.07 & 60.18 &  95.8 & 83.1 \\ & SecPE (Ciphertext) &  68.33 & 49.18 & 77.01 & 47.38 & 61.98 &  97.1 & 87.6

\\[0.5ex]\hline\\[-1.5ex]
\end{tabular}
\caption{Zero-shot accuracy on CIFAR-10, CIFAR-100, ImageNet and its variants.}
 \label{table:clip_benchmark}
 \vspace{-1em}
\end{table*}

Table \ref{table:clip_benchmark} shows the zero-shot accuracy on CIFAR-10, CIFAR-100, ImageNet, and its variants. Compared with plaintext inference, the ciphertext inference accuracy of {\sf SecPE} and Phoenix is slightly lower, which is caused by errors introduced by homomorphic encryption calculations. Experiments show that compared to Phoenix, {\sf SecPE}'s accuracy is higher. For example, in the setting of 247 prompts, {\sf SecPE}'s accuracy is 97.1\%, which is only 0.4\%  lower than the plaintext inference accuracy. However, Phoenix's accuracy is only 95.8\%, 1.7\% lower.

It can be demonstrated that each multiplication and rotation operation of RNS-CKKS will introduce a component of the error. The polynomial fitting of the sign operation necessitates a large number of multiplication calculations. Under the setting of 247 prompts, Phoenix requires 248 sign operations and ciphertext rotation, whereas {\sf SecPE} only requires 8 sign operations and ciphertext rotation, so the error of {\sf SecPE} will be significantly smaller.

\subsection{Efficiency Comparison}
\begin{figure}[htbp]
    \centering
    \includegraphics[width=3in]{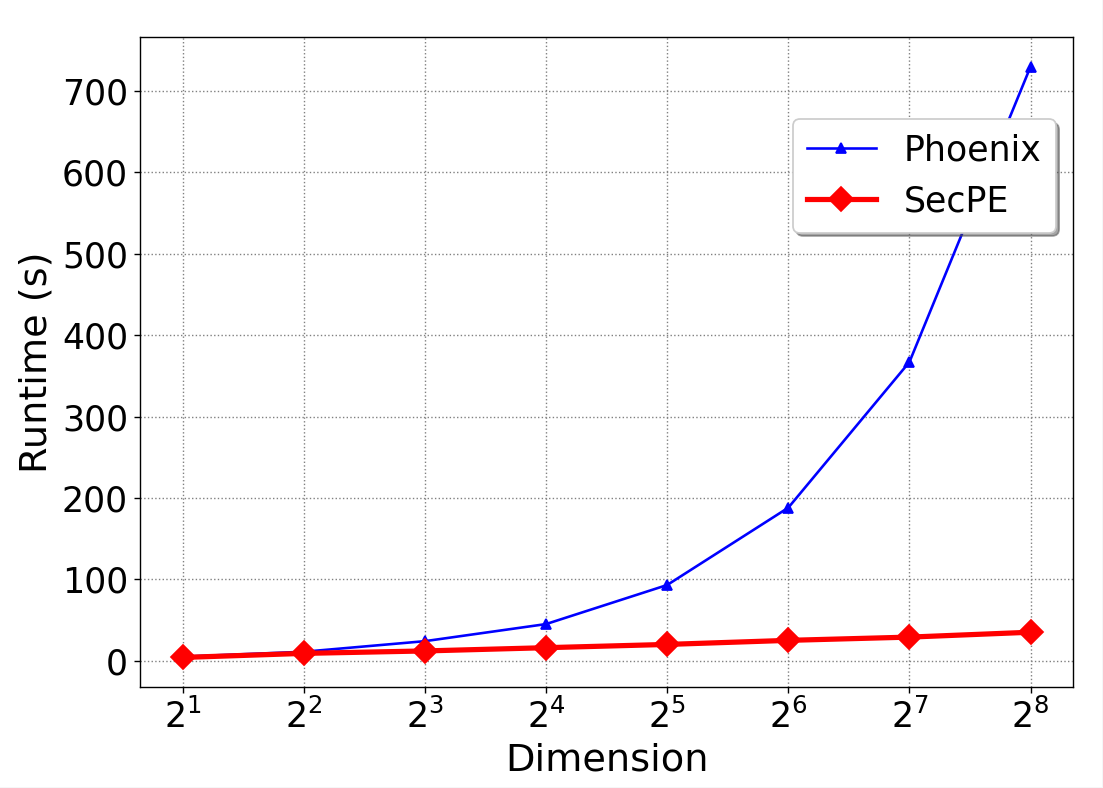}
    \caption{Performance of \textsc{Argmax} on RNS-CKKS for different dimensions of input.}
    \label{fig:evaluation}
    \vspace{-1em}
\end{figure}

Figure \ref{fig:evaluation} illustrates the efficiency comparison of \projectname ~with Phoenix \cite{jovanovic2022private} under different input dimensions.
In particular, we focus on the essential \textsc{Argmax} operation, which incurs one of the major overheads of prompt ensemble under private inference. For an input length of $n$, Phoenix \cite{jovanovic2022private} adopts a sequential comparison approach to obtain the sign bit, resulting in $(n+1)$ $Sign$ operations and $(n+1)$ ciphertext rotations. In contrast, \projectname's Algorithm \ref{alg:argmax} only requires $(\log n + 1)$ $Sign$ operations and $(\log n + 1)$ ciphertext rotations. This significantly reduces the execution time, which is depicted in Figure \ref{fig:evaluation}. For the input length of 256, \projectname ~achieves 20.8$\times$ speedup for \textsc{Argmax}. 

\begin{figure}[!h]
    \centering
    \includegraphics[width=3.3in]{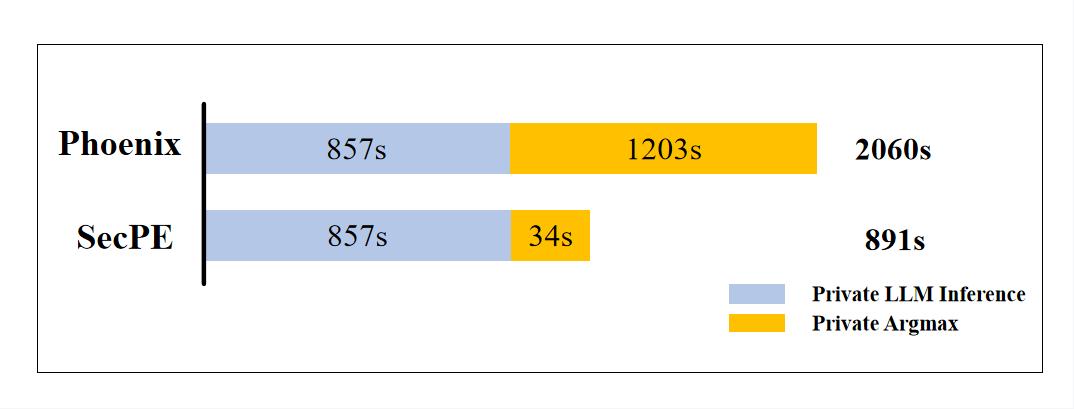}
    \caption{Comparison of Total Runtime and Argmax Computation Time between Phoenix and \projectname.}
    \label{fig:occupy}
\end{figure}

Figure \ref{fig:occupy} shows the time distribution of different building blocks in \projectname. Experiments show that $\textsc{Argmax}$ is an bottleneck of the prompt ensembling in ciphertext. The runtime overhead of argmax is even longer than the overhead of privacy LLM inference. \projectname~ significantly reduces the $\textsc{Argmax}$  computation time from 2060s to 891s, enhancing the overall efficiency. Therefore, \projectname ~incurs an additional cost of only $3.8\%$ compared to private inference with LLM without prompt ensembling.

It indicates that while Prompt Ensembling requires multiple inference runs, this overhead is justified. Despite the additional computational cost, as visualized by the substantial slice of the pie chart allocated to LLM inference, the benefits of Prompt Ensembling cannot be overstated. The improved robustness and accuracy provided by multiple inferences, where different prompts are evaluated to derive a final answer, results in more reliable and accurate model performance. This benefit often outweighs the cost of increased inference time, making prompt ensembling a valuable technique in scenarios where high-quality predictions are paramount.





\section{Conclusions}
We propose \projectname, the first attempt to our knowledge to jointly enable privacy-preserving and adversarial robustness for LLM inference. \projectname synergizes the strengths of private inference and prompt ensembling, previously studied in isolation, and overcomes the inefficiencies of a naive combination of existing techniques. 

Our extensive experiments have shown that \projectname~ not only maintains high clean accuracy but also significantly improves robustness, all with minimal efficiency overhead compared to existing private inference methods. Thus, \projectname~ manifests a satisfactory ``accuracy-robustness-efficiency'' tradeoff. The future work is to take advantage of existing hardware acceleration technology, such as GPU \cite{wang2023he} and FPGA \cite{viand2023heco}, which is expected to increase the efficiency by hundreds of times to achieve practical private inference.
\begin{ack}
This work was supported by the “Pioneer” and “Leading Goose” R\&D Program of Zhejiang (No. 2022C01126).
\end{ack}
\clearpage
\bibliography{mybibfile.bib}

\end{document}